\documentclass[
    ,final            
  ]
  {aipproc}

\layoutstyle{6x9}

\begin{document}

\title{Multiplicity Fluctuations in Heavy Ion Collisions at CERN SPS}

\classification{24.60.Ky}


\keywords {Multiplicity Fluctuations, NA49, CERN SPS}

\author{Benjamin Lungwitz\\
 for the NA49 Collaboration}{
  address={Fachbereich Physik der Universit\"{a}t, Frankfurt, Germany.}
}

\begin{abstract}
The system size and centrality dependence of multiplicity fluctuations in nuclear collisions 
at $158A$ GeV as well as the energy dependence for the most central
$Pb+Pb$ collisions were studied by the NA49 experiment at CERN SPS.
A strong increase of fluctuations was observed with decreasing centrality in $C+C$, $Si+Si$ and $Pb+Pb$ collisions.
The string hadronic models (UrQMD, Venus, HIJING, HSD) can not reproduce the observed increase. This may indicate a strong mixing
of target and projectile contribution in a broad rapidity range.
For the most central collisions at all SPS energies multiplicity distributions are significantly narrower than a corresponding Poisson one
both for negatively and positively charged hadrons.
The UrQMD model seems to reproduce the measured values on scaled variance.
Statistical model calculations overpredict results when conservation laws are not taken into account.

\end{abstract}

\maketitle


\section{Introduction}
At high energy densities ($\approx 1\, GeV/fm^3$) a phase transition from hadron gas to quark-gluon-plasma (QGP) is expected to occur. 
There are indications that at top SPS energies quark-gluon-plasma is created at the early stage of heavy ion collisions~\cite{Heinz:2000bk}.
Lattice QCD calculations suggest furthermore the existence of a critical point in the phase diagram of strongly interacting matter which separates the 
line of the first order phase transition from a crossover.
Models predict an increase of multiplicity fluctuations near the onset of deconfinement~\cite{Gazdzicki:2003bb} 
or the critical point~\cite{Stephanov:1999zu}.
Statistical model calculations~\cite{Begun:2004gs} showed a non-trivial decrease of fluctuations due to conservation laws. These reduced
fluctuations would then serve as a ``background'' in the search for the critical point.

\section{The NA49 Experiment}
The NA49 detector
is a large acceptance fixed target hadron spectrometer described in~\cite{Afanasev:1999iu}.
The centrality of a collision is determined using a downstream Veto calorimeter which measures the energy in the projectile spectator domain~\cite{Lungwitz:2006fp}.
This allows a determination of the number of projectile participant nucleons $N_P^{proj}$.
The number of target participants is not fixed in the experiment, model calculations show large fluctuations of their number in peripheral 
collisions~\cite{Konchakovski:2005hq}.

\subsection{Analysis Procedure}

Since detector effects like track reconstruction efficiency might have a large influence on multiplicity fluctuations, it is important to 
select a very clean track sample for the
analysis. Therefore the acceptance for this analysis is limited to a part of the forward hemisphere, where the NA49 detector has the
highest tracking efficiency.
This was done by restricting the analysis to the rapidity interval $1<y(\pi)<y_{beam}$
\footnote{Rapidity is calculated in the center of mass system assuming pion mass.}
 for $20A$ to $80A$ GeV and $1.08<y(\pi)<2.57$ for $158A$ GeV. In addition a cut on transverse momentum according 
to \cite{Alt:2004ir} was applied, which is dependent both on rapidity and azimuthal angle.
For such tracks the reconstruction efficiency is larger than $98\%$.

The basic measure of multiplicity fluctuations used in this analysis is the scaled variance:
$\omega=\frac{Var(n)}{<n>}$,
where $Var(n)$ and $<n>$ are variance and mean of the multiplicity distribution, respectively. In this paper only the results for
negatively charged hadrons are shown.

The data is corrected for the finite width of centrality bins. Data on centrality and system size dependence at $158A$ GeV is corrected for
the resolution of the Veto calorimeter, for most central $Pb+Pb$ data this correction is small (it would decrease
the results by less than $5\%$) and was neglected.

The total systematic error is estimated to be $+2\%$ respectively $-5\%$~\cite{Lungwitz:2006fp} for the most central $Pb+Pb$ collisions, 
the systematic error for $p+p$, $C+C$, $Si+Si$ and non-central $Pb+Pb$ collisions is shown in figure~\ref{centr_dep}.

\section{Centrality and System Size Dependence}
\begin{figure}
\includegraphics[width=16cm]{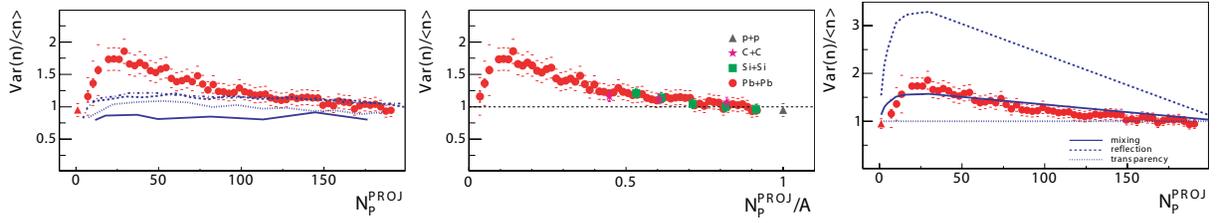}
\caption{\label{centr_dep}
Left: Comparison of scaled variance of $h^-$ in $Pb+Pb$ collisions (circles) with string hadronic models: HIJING~\cite{Gyulassy:1994ew}, 
UrQMD, HSD~\cite{Konchakovski:2005hq} and Venus~\cite{Werner:1993uh}.
Middle: Centrality dependence of scaled variance of $h^-$ in $p+p$, $C+C$, $Si+Si$ and $Pb+Pb$ collisions at $158A$ GeV.
Right: Scaled variance of $h^-$ in comparison to model calculations for transparency, mixing and reflection of matter in the early 
stage~\cite{Gazdzicki:2005rr}.
The outer errors correspond to a sum of statistical and systematical uncertainties.}
\end{figure}

The centrality dependence of the scaled variance for negatively charged hadrons ($h^-$) at $158A$ GeV is shown in figure~\ref{centr_dep}. 
Scaled variance increases with 
decreasing centrality of a collision. For very peripheral collisions there is a hint that it decreases again, but the systematic errors are large in this 
region.
Data is compared to predictions of various string hadronic models for $Pb+Pb$ collisions; they all predict, in contradiction to data,
a flat centrality dependence.
Scaled variance behaves similar in $p+p$, $C+C$, $Si+Si$ and
$Pb+Pb$ collisions if plotted against centrality defined as $N_P^{proj}/A$, where $A$ is the number of nucleons in nuclei. 
This is a hint that not
the size of the collision system is the correct scaling parameter for the observed increase of $\omega$ but the 
fraction of the colliding nucleons in the projectile nuclei.
In \cite{Gazdzicki:2005rr} it is suggested that target participants contribute to particle production in the projectile hemisphere and 
their fluctuations therefore result in an increase of
multiplicity fluctuations. The predictions of this effect called mixing is in approximate agreement with data as shown in figure~\ref{centr_dep} (right).
Inter-particle correlations~\cite{Rybczynski:2004zi} and percolation models~\cite{Cunqueiro:2005hx} provide alternative explanations of the results.

\section{Energy Dependence in Central Collisions}

\begin{figure}
\includegraphics[height=6cm]{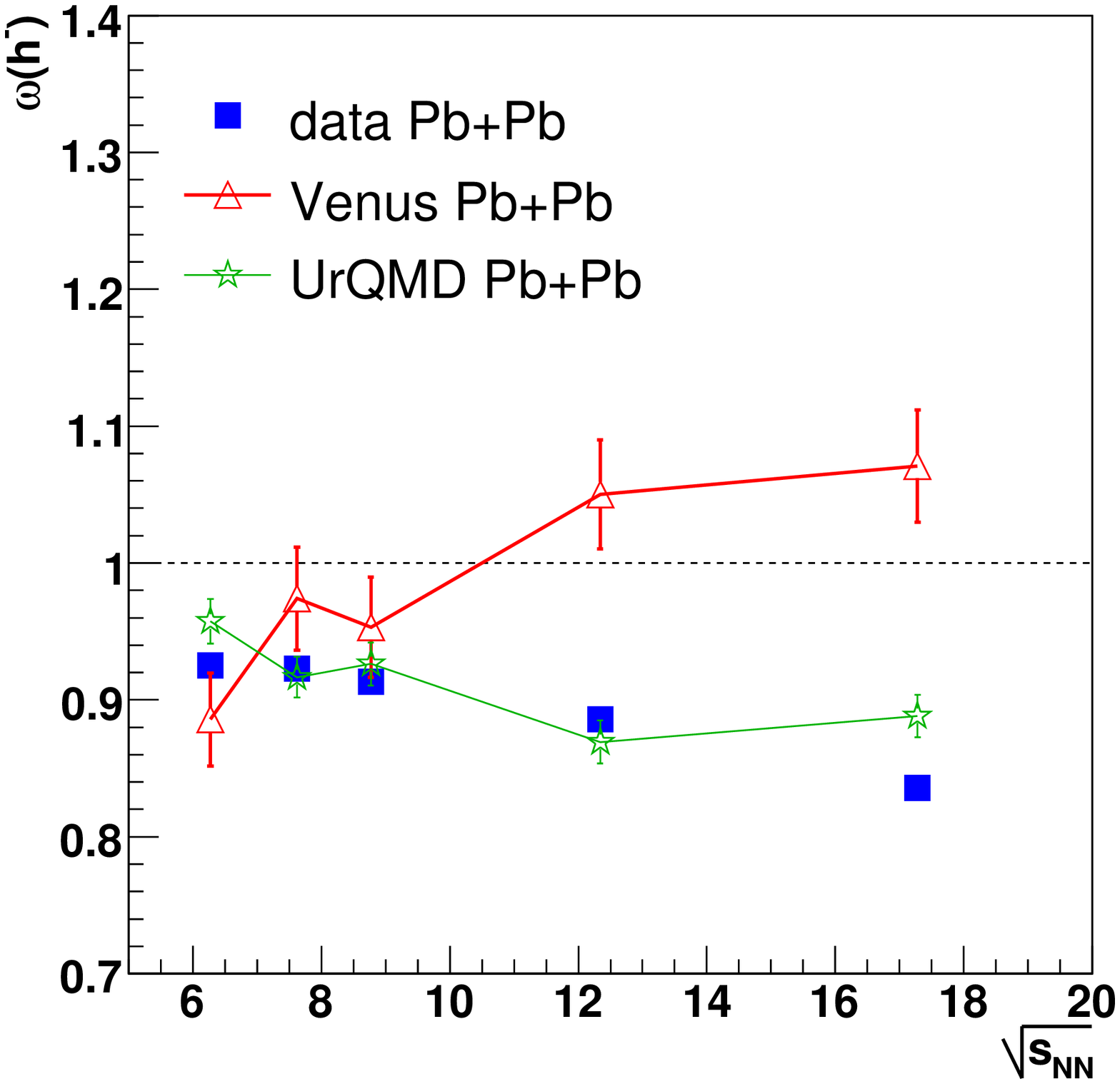}
\includegraphics[height=6cm]{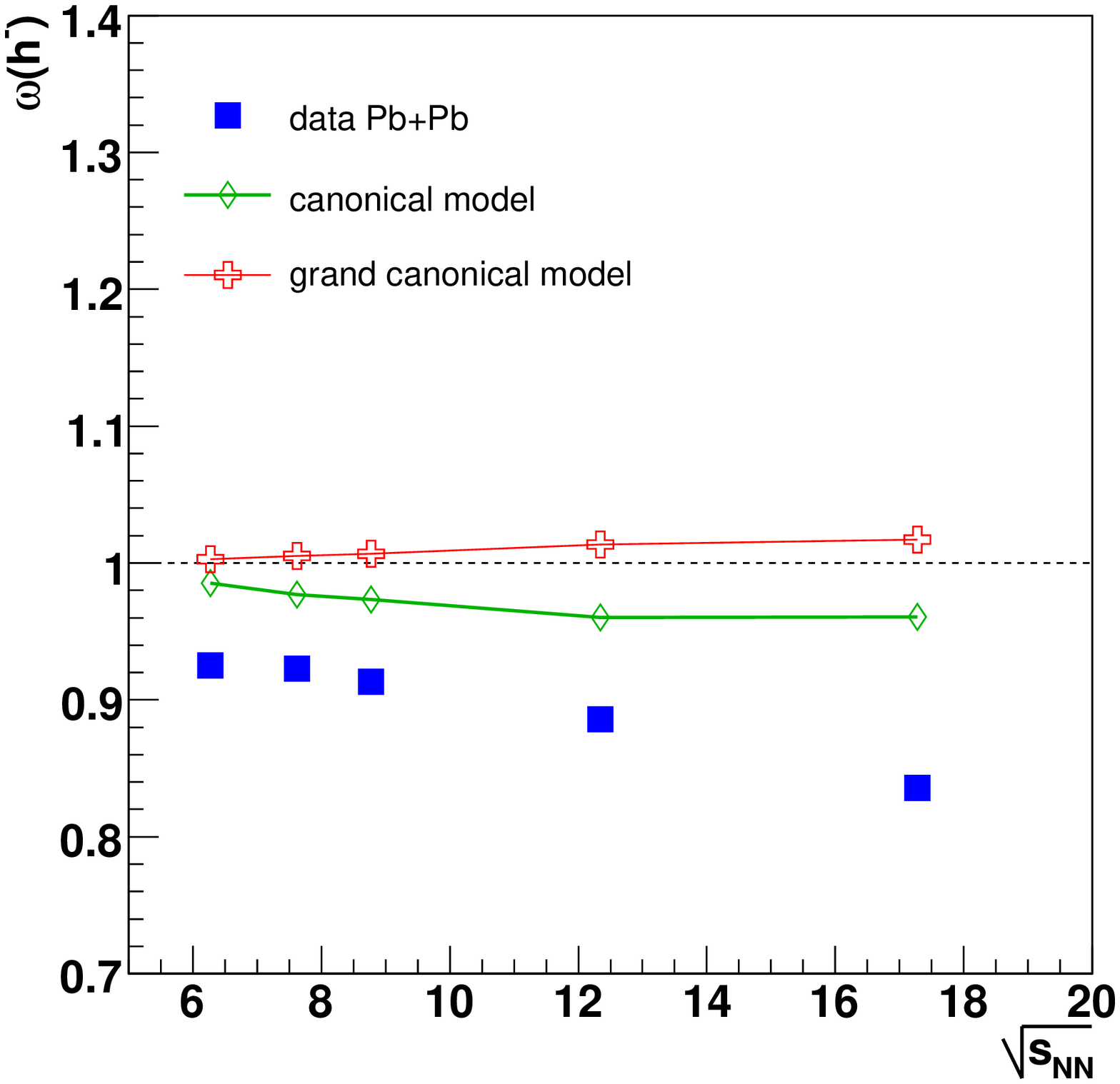}
\caption{\label{ed_mod} Energy dependence of multiplicity fluctuations of $h^-$ in $Pb+Pb$ collisions in comparison to string-hadronic 
models Venus~\cite{Werner:1993uh} and UrQMD~\cite{Bass:1998ca} (left) and in comparison to 
canonical and grand-canonical statistical hadron-resonance gas models~\cite{Begun:2006jf} (right). Only statistical errors are shown, the systematical
errors (not shown) are $\sigma_{sys}\approx^{+2}_{-5}\%$.}
\end{figure}
At all energies the scaled variance of $h^-$ is significantly smaller than one, the value for a corresponding Poisson
distribution (see figure~\ref{ed_mod})~\cite{Lungwitz:2006fp}. 
A direct quantitative comparison of $\omega$ at different energies is not possible due to different experimental
acceptance.
The UrQMD model is in approximate agreement with data. The Venus model overpredicts $\omega$ at higher energies.
No significant increase of multiplicity fluctuations due to critical point or onset of 
deconfinement is observed.

Predictions of a statistical model~\cite{Begun:2006jf} for canonical and grand-canonical ensembles are compared to data.
The measured scaled variance is much lower than predicted by the grand-canonical ensemble. The canonical model predicts $\omega$ smaller than $1$, in 
qualitative agreement with data.
Energy momentum conservation and the finite volume of hadrons are expected to cause an additional suppression of fluctuations.
Thus the observed small fluctuations seem to be a non-trivial effect of conservation laws in a relativistic hadron gas~\cite{Begun:2004gs}.




\bibliographystyle{aipproc}   

\bibliography{proceedings}

\IfFileExists{\jobname.bbl}{}
 {\typeout{}
  \typeout{******************************************}
  \typeout{** Please run "bibtex \jobname" to optain}
  \typeout{** the bibliography and then re-run LaTeX}
  \typeout{** twice to fix the references!}
  \typeout{******************************************}
  \typeout{}
 }

\end{document}